\documentclass[pdflatex,sn-mathphys-num,iicol]{sn-jnl}

\usepackage{graphicx}%
\usepackage{multirow}%
\usepackage{amsmath,amssymb,amsfonts}%
\usepackage{amsthm}%
\usepackage{mathrsfs}%
\usepackage[title]{appendix}%
\usepackage{xcolor}%
\usepackage{textcomp}%
\usepackage{manyfoot}%
\usepackage{booktabs}%
\usepackage[linesnumbered,ruled,vlined]{algorithm2e}
\usepackage{algorithmicx}%
\usepackage{algpseudocode}%
\usepackage{listings}%

\theoremstyle{thmstyleone}%
\theoremstyle{thmstyletwo}%

\theoremstyle{thmstylethree}%

\begin{document}

\title[Article Title]{False Positives Raised by Quantum Readout Error Mitigation}

\author[1,5,6]{\fnm{Yibin} \sur{Guo}}\equalcont{These authors contributed equally to this work.}

\author[2]{\fnm{Yi} \sur{Fan}}\equalcont{These authors contributed equally to this work.}

\author[1]{\fnm{Pei} \sur{Liu}}\equalcont{These authors contributed equally to this work.}

\author[1]{\fnm{Shoukuan} \sur{Zhao}}

\author*[3]{\fnm{Xiaoxia} \sur{Cai}}\email{xxcai@ihep.ac.cn}

\author*[2]{\fnm{Xiongzhi} \sur{Zeng}}\email{xzzeng@ustc.edu.cn}

\author*[2]{\fnm{Zhenyu} \sur{Li}}\email{zyli@ustc.edu.cn}

\author*[1]{\fnm{Wengang} \sur{Zhang}}\email{zhangwg@baqis.ac.cn}

\author[1]{\fnm{Yirong} \sur{Jin}}

\author[1,4]{\fnm{Hai-Feng} \sur{Yu}}

\affil*[1]{\orgdiv{}, \orgname{Beijing Academy of Quantum Information Sciences}, \orgaddress{\street{}, \city{Beijing}, \postcode{100193}, \state{}, \country{China}}}
\affil[2]{\orgdiv{State Key Laboratory of Precision and Intelligent Chemistry}, \orgname{University of Science and Technology of China}, \orgaddress{\street{96 Jinzhai Road}, \city{Hefei}, \postcode{230026}, \state{Anhui}, \country{China}}}
\affil[3]{\orgdiv{Institute of High Energy Physics}, \orgname{Chinese Academy of Sciences}, \orgaddress{\street{}, \city{Beijing}, \postcode{100049}, \state{}, \country{China}}}
\affil[4]{\orgdiv{}, \orgname{Hefei National Laboratory}, \orgaddress{\street{}, \city{Hefei}, \postcode{230088}, \state{}, \country{China}}}
\affil[5]{\orgdiv{Institute of Physics}, \orgname{Chinese Academy of Sciences}, \orgaddress{\street{}, \city{Beijing}, \postcode{100190}, \state{}, \country{China}}}
\affil[6]{\orgdiv{}, \orgname{University of Chinese Academy of Sciences}, \orgaddress{\street{}, \city{Beijing}, \postcode{101408}, \state{}, \country{China}}}


\abstract{
Quantum readout error mitigation is essential for noisy intermediate-scale quantum devices to achieve reliable data. 
The conventional approaches, conflating initialization errors with measurement errors, not only suppress the influence of measurement errors, but also strengthen that of initialization errors, which is a systematic bias grows exponentially with the qubit number.
Here, we have proved that this effect causes severe fidelity overestimation for all stabilizer states and might lead to false positives in large-scale entangled state characterization.
Similarly, the results from algorithms like the variational quantum eigensolver and time evolution also deviate negatively, and cover up other errors in the quantum circuit.
These findings highlight the critical need for rigorous benchmarking and careful management of initialization errors.
Consequently, we establish an upper bound for the tolerable initialization error rate to ensure effective error mitigation at a given system scale.}

\maketitle

\section*{Introduction}\label{sec1}

In the ongoing noisy intermediate-scale quantum (NISQ) era \cite{bharti2022noisy}, quantum error mitigation (QEM) advanced through various methods in recent years is essential for extracting valuable information from noisy devices by suppressing diverse error sources \cite{temme2017error, berg2023probabilistic, cai2023QEM}.
Among the QEM methods, quantum readout error mitigation (QREM) is a critical step toward achieving accurate results and is usually realized in practice by profiling the measurement error matrix and computing its inverse \cite{shen_correcting_2012,zhong_violating_2019, maciejewski_mitigation_2020, nation_scalable_2021,mooney_generation_2021, cao_generation_2023, noauthor_experimental_2024, song_realization_2025,jiang2025generation,javadi2025big}.
When calibrating the measurement error matrix, state preparation errors are inevitably conflated with measurement errors, since the state preparation and measurement (SPAM) errors are considered inherently difficult to separate.
However, this approach is generally regarded to be valid under the assumption that state preparation errors (also known as the initialization error) are negligible compared to the measurement errors (also known as the readout error).
For instance, in current superconducting transmon qubit systems, state initialization error rates are typically one to two orders of magnitude lower than measurement error rates, and are therefore routinely neglected in experiments.
Consequently, the conventional QREM has been regarded as safe for application \cite{javadi2024quantum}.

Nevertheless, in this work, we show that even when state initialization errors are orders of magnitude smaller than measurement errors, the conventionally neglected state preparation component induces a systematic bias in corrected observable expectation values that scales exponentially with the number of qubits.
This effect, which is negligible on small-scale processors and typically overlooked, becomes significant and requires careful consideration as system size increases.
We first show that the mixture of SPAM error leads to severe overestimation in large-scale state fidelity benchmarking with QREM.
Furthermore, we demonstrate that these widely used quantum algorithms reliant on observable estimation, such as the variational quantum eigensolver (VQE) \cite{Tilly_VQE_2021, Cerezo2021, CaoRomOls19, PerMcCSha14, Kandala2017, LanWhi10, mcardle2020quantum,noauthor_experimental_2024} and the quantum time evolution (QTE) \cite{Babbush_chem_trotter_2015, Hastings_qalgo_chem_2015, Jones_opt_trotter_2019, Avtandilyan_opt_trotter_2024, sarkar_scalable_qte_2024, Tranter_trotter_err_2019, Whitfield_qchem_2011, Yang_high_order_trotter_2022}, are also adversely affected.
Finally, for future applications, we derive an upper bound on initialization errors as a function of the qubit number.
By satisfying this bound, the deviation in outcomes remains tightly controlled, thereby ensuring reliable and trustworthy results from quantum information processing.

\section*{Results}\label{sec2}

\subsection*{Introduction to QREM}
For an $n$-qubit system, the experimentally measured probability distribution can be treated classically and satisfy
\begin{equation}
    {\boldsymbol{P}}_{\text{noisy}}=M {\boldsymbol{P}}_{\text{ideal}},
\end{equation}
where ${\boldsymbol{P}}_{\text{noisy}}$ (${\boldsymbol{P}}_{\text{ideal}}$) is a vector containing $2^n$ outcome probabilities with (without) measurement error and $M$ is a $2^n \times 2^n$ matrix characterizing the most general measurement error.
To characterize $M$, the qubits are selectively flipped and the noisy outcomes are measured under all $2^n$ initial states.
Specially, if the measurement error correlations between different qubits are not considered, the measurement error matrix $M$ takes the form $M=\otimes_i M_i$, where $M_i$ is the measurement error matrix of each qubit.
Meanwhile, the initialization errors can also be modeled as independent and incoherent errors on individual qubits.
Then, for each qubit, we use an error mitigation matrix $\Lambda_i$ satisfying
\begin{equation}
    \begin{pmatrix}
        1&0\\0&1
    \end{pmatrix}=\Lambda_i M_i \begin{pmatrix}
        1-q_i &q_i \\ q_i  & 1-q_i
    \end{pmatrix},
\end{equation}
to mitigate the measurement error with the initialization error rate $q_i$ on qubit $Q_i$.
For the whole system, the mitigation matrix $\Lambda$ is also the tensor form of the individual correction matrix $\Lambda_i$.

It is noticed that, the mitigation matrix $\Lambda$ is not simply the inverse of measurement error matrix $M$, but also affected by the initialization error of the qubits.
When the initialization error rate is indeed much smaller than the readout error rate, the difference between $\Lambda$ and $M^{-1}$ is generally considered negligible.
However, as qubit number grows, this error will accumulate and cause large deviation onto the observable measurement.

\subsection*{Effects on large-scale stabilizer state characterization}

\begin{figure*}
\centering 
\includegraphics[]{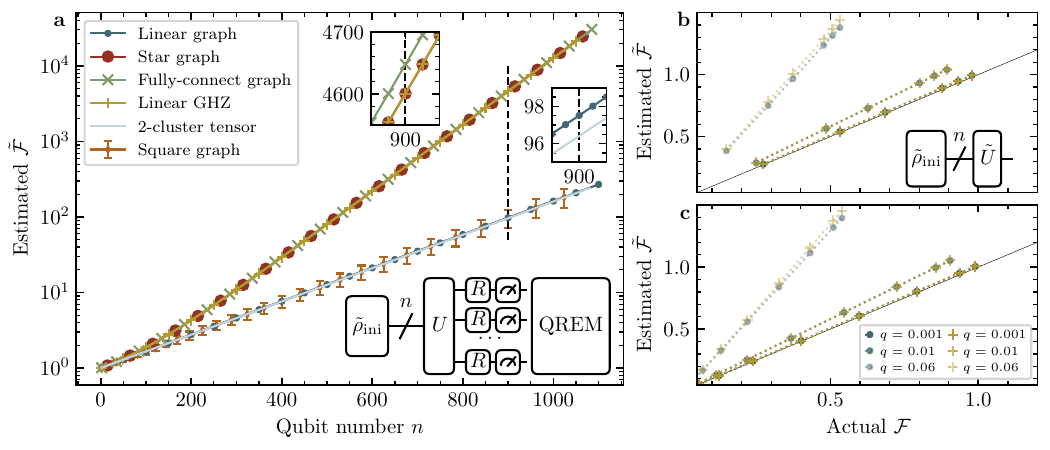}
\caption{\textbf{Fidelity overestimation of large-scale entangled states}. \textbf{a}, The fidelity overestimation of different entangling states, including the linear graph state, the star-liked graph state, the fully-connected graph state, the linear GHZ state, and the squared-grid graph state.
The data of square graph with its average and its standard deviation is calculated by random sampling methods, and the others are exact solutions by dynamic programming.
Here, only SPAM error is considered ($\tilde\rho_i$) and the initialization error rate is set to be $q=0.01$.
The overestimation of the $10$-qubit linear graph state (labeled by blue dots) and the $10$-qubit linear GHZ state (labeled by green crosses) are illustrated with depolarization error of controlled-Z (CZ) gates in \textbf{b} and with controlled-rotation error in \textbf{c}, respectively.}
\label{entangle_fig}
\end{figure*}
We first consider the influence of QREM on large-scale state fidelity benchmarking.

For an $n$-qubit stabilizer state $|\varphi\rangle$, the fidelity of the experimentally prepared state is defined as $\mathcal F=\textrm{Tr}(\rho_{\text{exp}} \rho_{\text{ideal}})$, where $\rho_{\text{ideal}}=|\varphi \rangle \langle \varphi |$ is the density matrix of the perfect entangled state and $\rho_{\text{exp}}$ is the experimentally prepared density matrix.
For small scale systems, the density matrix is usually obtained via full quantum state tomography (QST) which takes at least $3^n$ measurements.
An alternative method \cite{cao_generation_2023, jiang2025generation,chen_nishimori_2025,javadi2025big} leverages the stabilizer structure of the entangled state,
\begin{equation}
    \rho_{\text{ideal}}=\underset{i}{\prod} \frac{I+g_i}{2}=\frac1{2^n}\sum_kS_k,
\end{equation}
where $g_i$ is the stabilizer generator and $S_k$ is the stabilizer of the state,
i.e. $S_k\in \mathcal G=\langle g_0,g_1,...,g_{n-1}\rangle$.
The average measurement outcome of the stabilizers $S_k$,
\begin{equation}
    \mathcal F=\frac1{2^n}\sum_{k=0}^{2^n-1}\mathrm{Tr}(\rho_{\text{exp}}S_k),
\end{equation}
gives the state fidelity $\mathcal F$.
It requires only $2^n$ measurements, less than the $3^n$ of QST but still scaling exponentially. Thus, the randomized sampling method is developed whose overhead does not actually increase with the qubit number \cite{flammia_direct_2011}.
The average measurement outcome of these randomly chosen stabilizers gives an unbiased estimation of the state fidelity which allows us to benchmark large scale entangled state fidelity with high efficiency.

When initialization error is considered, the final density matrix can be written in the stabilizer form as:
\begin{equation}
    \rho_{\text{noisy}}=\underset{i}{\prod} \frac{I+(1-2q_i)g_i}{2},
\end{equation}
where we tentatively assume that the entangled state preparation process is perfect and $q_i$ is the initialization error rate on $Q_i$.
We use a stabilizer $S$ which consists of two stabilizer generator $g_0=X_0I_1X_2Z_3I_4I_5$ and $g_2=I_0Y_1I_2Z_3I_4Y_5$ ($S=g_0g_2=X_0Y_1X_2I_3I_4Y_5$) as an example to illustrate the origin of the overestimation.
Ideally, when the measurement is perfect, the measured expectation value would be $\langle S\rangle=\mathrm{Tr}(\rho_{\text{noisy}}S)=(1-2q_0)(1-2q_2)$.
Thus, according to ref \cite{van_den_berg_model-free_2022} and the derivations in Methods, the expectation value after QREM correction is $\langle S\rangle_{\rm QREM}=(1-2q_1)^{-1}(1-2q_5)^{-1}>1$.

For arbitrary $n$-qubit system, $\forall i$, there always exists at least one stabilizer generator $g_j$, such that the $g_j$ does not contain an $I$ on the $i$th qubit.
Otherwise, the $i$th qubit is not stablized by any stabilizer, and the system still has degrees of freedom \cite{nielsen2010quantum}.
We consider a subgroup of $\mathcal G$ defined as  ${\mathcal{G}_j}=\langle g_0,g_1,...,g_{j-1},g_{j+1},...,g_{n-1}\rangle$ and its coset ${g_j\mathcal{G}_j}$ with $g_j\mathcal{G}_j\cup\mathcal G_j=\mathcal G$.
For $\forall S_m \in \mathcal{G}_j$, there exists a unique stabilizer $S_{m'}\in g_j\mathcal{G}_j$ which satisfies  $S_{m'} = g_j S_m$.
Since $g_j$ does not contain an $I$ on the $i$th qubit, $S_m$ and $S_{m'}$ do not contain two $I$s on the $i$th qubit at the same time.
Then $\forall S_m,S_l\in\mathcal{G}_j$, if $S_m\neq S_l$,
$S_{m'}$ is not equal to $S_{l'}$.
Thus $\forall i$, the fraction $f_{i,I}$, defined as the ratio of stabilizers containing $I$ on the $i$th qubit to the total number of stabilizers, satisfies $f_{i,I} \leq 1/2$.
The state fidelity $\tilde{\mathcal F}$ is then estimated to be
\begin{equation}
\begin{split}
\tilde{\mathcal F}&=\frac{1}{2^n} \sum_{j=0}^{2^n-1} \langle S_{j} \rangle_{\rm QREM} \geq \sqrt[2^n]{\prod_{j=0}^{2^n-1} \langle S_j\rangle _{\rm QREM}}\\
&=\sqrt[2^n]{\prod_{i=0}^{n-1} (1-2q_i)^{2^{n-1}-2^{n}(1-f_{i,\mathrm{I}})}} \geq 1,
\end{split}
\label{eq:fidelity}
\end{equation}
where we have used the basic inequality $\sum{a_i} /n \geq \sqrt[n]{\prod a_i}$.
In Methods, we proved that the equalities in Eq.\ref{eq:fidelity} hold \textbf{only if} the system is prepared in a tensor product of single qubit states.
Therefore, for all stabilizer states, if entanglement exists, the state fidelity achieved through stabilizer expectation value measurement is overestimated after the conventional QREM correction.

Here, we use several typical graph states (GSs) and the Greenberger-Horne-Zeilinger (GHZ) state as examples to quantitatively demonstrate the significant fidelity over estimation.
To precisely quantify the fidelity overestimation while eliminating sampling uncertainty, we compute the expectation values of all stabilizers rather than the randomly picked stabilizers (see Methods for details).
Figure~\ref{entangle_fig}~\textbf{a} presents the test circuit and the final state fidelity achieved via the conventional QREM method, termed "estimated fidelity" $\tilde{\mathcal{F}}$ hereafter.
Notably, the overestimation of estimated fidelity grows exponentially-like with the size of the entangled state, including the linear graph state, the star-liked graph state, the fully-connected graph state, and the linear GHZ state \cite{hein2006entanglement}.
In addition, we also consider a state we called the "2-cluster tensor" which is the tensor product of $n/2$ pairs of two-qubit cluster state.
Despite the slower growth of its estimated fidelity, we prove mathematically that the growth of this fidelity is exponential (see Methods for details).
For different entangled states, the growth of the estimated fidelity exhibits distinct behaviors, arising from differences in their preparation circuits that cause initialization errors to propagate in different patterns (see Methods for details).

The fidelity overestimation with imperfect entanglement preparation process is shown in FIG.~\ref{entangle_fig}~\textbf{b} and \ref{entangle_fig}~\textbf{c}.
The preparation process consists of single qubit gates and CZ gates.
We consider two kind of errors on the CZ gates, the depolarization error in Fig.\ref{entangle_fig}~ \textbf{b} and the control phase deviation in Fig.\ref{entangle_fig}~\textbf{c}.
In both cases, fidelity after QREM seems approximately overestimated by a certain factor.
Particularly, the estimated fidelity $\tilde{\mathcal F}$ could reach $0.5$ even if the actual fidelity $\mathcal F$ is not reached, leading to a false positive result in entanglement characterization.

\subsection*{Error induced by QREM in algorithm applications}
The VQE and QTE are important quantum algorithms for quantum chemistry applications which have the potential to demonstrate quantum advantage. While classical methods struggle with exponential scaling for many-body systems, quantum circuits offer an efficient path for encoding Hamiltonian $H$ or wavefunctions $|\Psi\rangle$. 

Both the VQE and QTE suffer from quantum errors but show distinct characteristics. Previous studies reveal that VQE exhibits partial resilience to coherent errors through parameter optimization \cite{vqe_noise_resilience, Kandala2017}.
However, whether this resilience extends to initialization errors is unclear.
Similarly, although Trotterization errors in QTE can be systematically overcome by introducing higher Trotter steps, additional deviations can arise from QREM-induced bias when initialization is imperfect.
Using simulation settings in Methods, we perform VQE with unitary coupled-cluster ansatz truncated at single and double excitations (UCCSD) and QTE for molecular systems to study the influence of QREM-induced errors.

\begin{figure}
\centering
\includegraphics[]{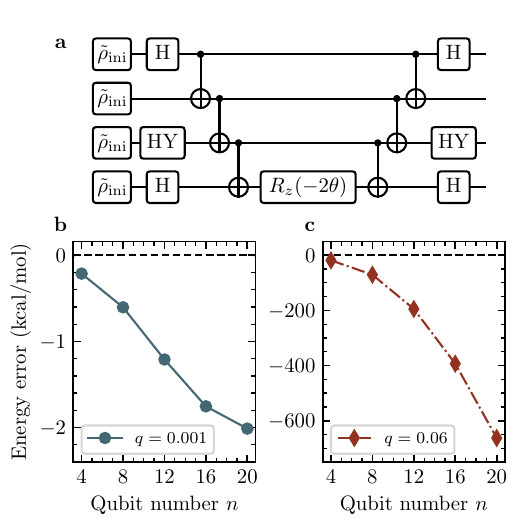}
\caption{\textbf{VQE simulation results for hydrogen chain}. \textbf{a}, $4$-qubit UCCSD circuit demonstration. Optimized ground state energy error with the initialization error rate \textbf{b} $q=0.001$ and \textbf{c} $q=0.06$ for H$_{2}$ (4 qubits) to H$_{10}$ (20 qubits).}
\label{vqe_gs}
\end{figure}

\begin{figure}
\centering
\includegraphics[]{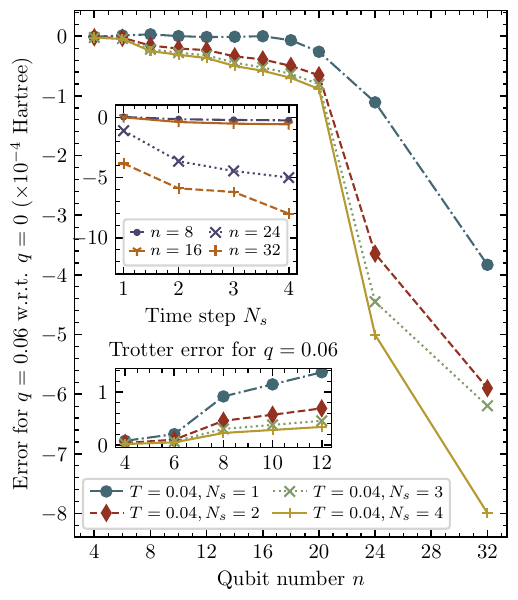}
\caption{\textbf{Time evolution errors under initialization error and conventional QREM}. Lower subplot shows Trotter error at initialization error $q=0.06$. Upper subplot displays time evolution errors with respect to $N_s$ for varying qubit number $n$.}
\label{fig:time_evo_fix_total_t}
\end{figure}

As demonstrated in FIG.~\ref{vqe_gs}, the VQE benchmarks for one-dimensional hydrogen chains reveal critical dependence on initialization error rate and qubit number.
Such a configuration lead to the number of qubits being twice of the number of atoms, $n = 2\times n_{\text{atom}}$.
With $q=0.001$ per qubit,
absolute energy error shows approximately linear scaling with qubit number under successful parameter optimization.
At $q = 0.06$, energy error exhibits accelerated scaling surpassing linear growth, indicating that QREM-induced errors become dominant and overwhelm ansatz's error-resilience at large $q$.

The results for QTE are shown in FIG.~\ref{fig:time_evo_fix_total_t}.
Given total time $T_{s}$ and number of Trotter steps $N_{s}$, the Trotter error $\Delta E_{\text{Trotter}}$ is defined as $\Delta E_{\text{Trotter}} = E_{T,N_{s};q} - E_{T, N_{s}=\infty;q}$,
and the energy error $\Delta E$ is calculated as $\Delta E = E_{T,N_{s};q} - E_{T, N_{s};q=0}$,
where $E_{T,N_{s};q}$ represents QREM-calibrated energy under initialization error rate $q$.
As expected, the Trotter error decreases with increasing $N_s$ when qubit count is fixed.
However, the total energy error exhibits divergent behavior where its magnitude grows with both qubit count and $N_s$, and finally surpasses Trotter error by a significant amount in large-qubit regimes (e. g., 32 qubits, $N_s = 4$).
This is consistent with VQE benchmarks, where QREM also severely breaks fidelity for ground-state optimizations.

\begin{figure}[!t]
    \centering
    \includegraphics[]{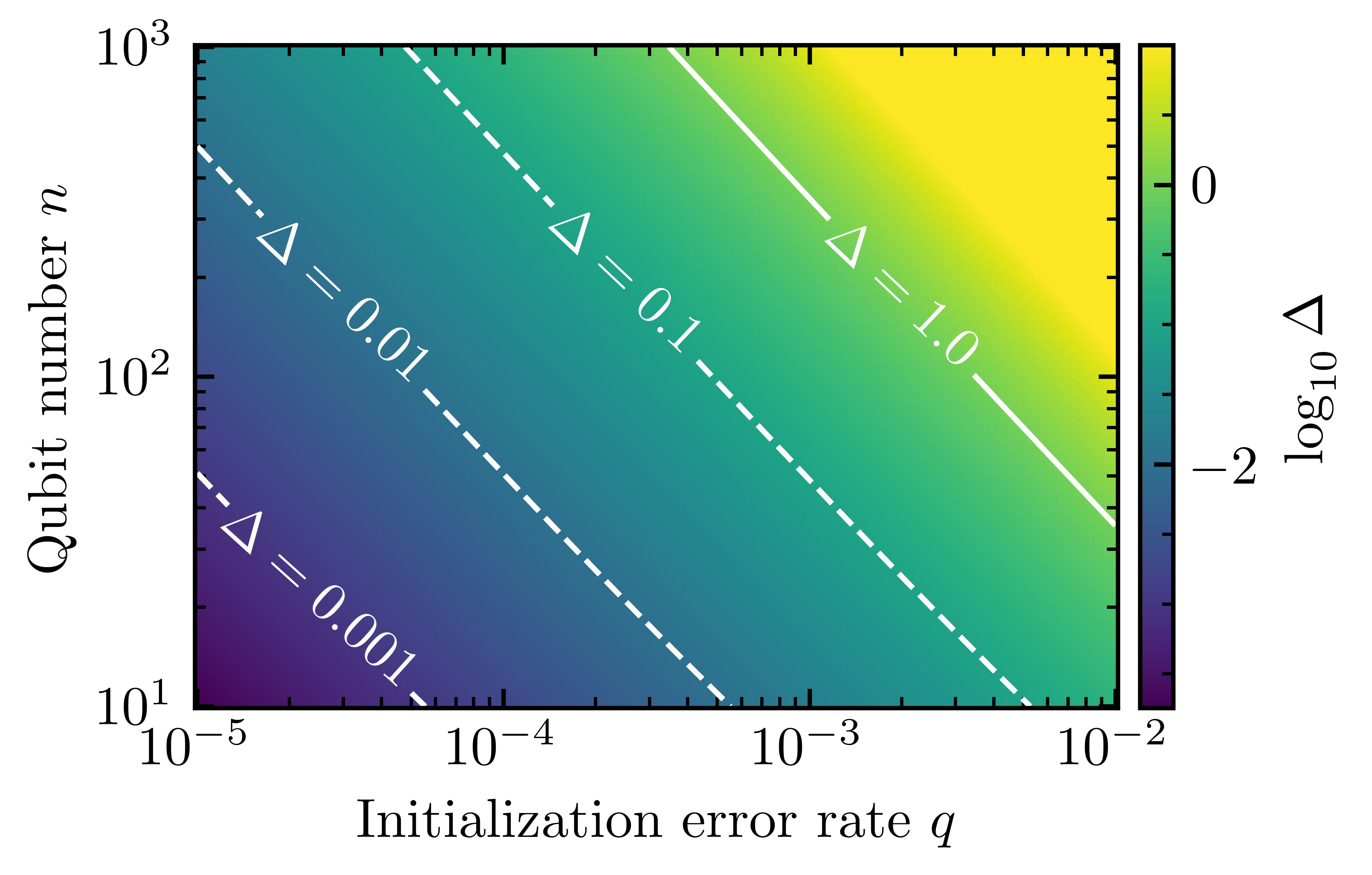}
    \caption{\textbf{The QREM induced error as a function of qubit number and initialization error.} Different lines delineate the upper bound on initialization error for a given system size. Only when system parameters fall below these lines can the error in the resulting outcome be rigorously bounded.}
    \label{fig:bound}
\end{figure}

\subsection*{Safety bounds for QREM}
Given that conventional QREM introduces systematic bias, it is imperative to strictly bound the initialization error rate for reliable results.
The calculation of the initialization error propagation through a quantum circuit is almost as difficult as the quantum circuit, especially with the non-Clifford process.
Therefore, we have to consider the worst-case scenario, the observable consists of only one stabilizer generator and involves all the qubits. 
From the analysis above, this kind of observable is mostly deviated from the real value.
In such case, the measured outcome is amplified by $(1-2q)^{-n+1}$.
It implies that, to keep QREM-induced errors within acceptable bounds, initialization errors must be suppressed more stringently as the number of qubits increases.
Figure~\ref{fig:bound} shows the relative error $\Delta =(\langle S \rangle_{\rm QREM}-\langle S \rangle)/\langle S \rangle$,
as a function of the initialization error rate $q$ and the qubit number $n$.
Without assuming uniform initialization errors, we perform a first-order Taylor expansion of the error, yielding $\Delta  \approx 2n \overline{q}$ where $\overline{q}$ is the average initialization error across all qubits. This provides an approximate upper bound on the tolerable error.
Since it is derived under the worst-case scenario, it provides rigorous safety guarantees for all expectation-value estimation experiments.
In recent quantum computing research, the characterization of the initialization error rate is limited by the error of readout and single-qubit operation.
Under current readout and single-qubit gate error conditions, an experiment that fails to satisfy this bound will exhibit unanticipated uncertainties.

\section*{Discussion}

In work, we point out that the conventional QREM method introduces a systematic error on operator measurement, which causes significant deviation on many kinds of experiments including entangling state characterizations, quantum algorithm applications, etc.
As we mentioned and proved, this deviation happens because of the initialization errors are propagated by the entangling gates in the quantum circuit. Therefore, any of the QREM methods which do not containing entangling gates properly during the calibration process can not get rid of this deviation.
Then, we have proved that for all stabilizer states, the fidelity would be overestimated.
Several entangling states are used to show that  the deviation grows exponentially with the system size.
We emphasize that this is important because the QREM-induced error might cover up the gate operation errors and give out false positive conclusions.
The extent of overestimation is determined by the neglected or unreported initialization error rate.
Moreover, the preparation of large-scale entangling states is widely used for performance benchmarking on quantum computing system.
However, here we proved that higher fidelity results is not only achieved by more accurate control of the system, but also can benefit from larger initialization error.
Thus, we suggest that the careful bounding of the initialization error is necessary.
Otherwise, this serious overestimation of fidelity will cause wrong judgment on performance benchmarking and hinder the development of quantum computing.
We also take some of the widely used quantum chemistry algorithms as examples to demonstrate that this effect also cause large deviation on real applications of quantum computation.
These applications indicate that the deviation of expectation value measurement also happens on non-Clifford circuits.
To implement full-scale quantum error correction remains to be an challenging task in a short term because of many difficulties.
For example, to prepare a 100-logical-qubit entangling state, tens of thousand physical qubits are required to suppressed the initialization error down to the safety bound we derived above.
Therefore, we suggest that when the system scale keeps increasing, more precise qubit reset techniques needs to be developed \cite{murch2012cavity,jader2025drift}.
Otherwise the quantum advantage based on operator expectation value measurement needs to be re-evaluated \cite{lanes2025framework}.
Meanwhile, for near-term applications, the self-consistent characterization and mitigation methods would be a possible solution \cite{edward2025disambiguating}.

\section*{Declarations}

\begin{itemize}

\item \textbf{Note added}:
We noted that during the preparation of this manuscript, there is another work pointed out that the state preparation error leads to a biased estimation of measured observable expectation values \cite{khan2025separate}. They also mentioned that under assumption of perfect single-qubit operation and perfect Quantum Non-Demolition (QND) measurement, the initialization error and measurement error can be separated from each other.

\item \textbf{Acknowledgments}:
We thank Dr. Huili Zhang, Dr. Huikai Xu, Dr. Kehuan Linghu, Prof. Xiao Yuan, and Prof. Liang Jiang for their helpful discussions. We thank the support from the National Natural Science Foundation of China (No. 22303005, No. 92365206, No. 12404557, No. 22393913, No. 22303090, and No. 12504576), the Strategic Priority Research Program (XDB0450101), the robotic AI-Scientist platform of the Chinese Academy of Sciences, and the Innovation Program for Quantum Science and Technology (No. 2021ZD0301802, No. 2023ZD0300200, and No. 2024ZD0301500).

\item \textbf{Competing interests}:
The authors declare that the authors have no competing interests as defined by Nature Portfolio, or other interests that might be perceived to influence the results and/or discussion reported in this paper.

\item \textbf{Data availability}:
We have provide all the simulation and calculation methods. All the data can be generate easily through the Methods section.

\item \textbf{Code availability}:
The code that support the findings of this study are available from the corresponding author upon reasonable request.

\item \textbf{Author contribution}:
S.Z. and W.Z. gave out the original idea.
Y.F. and X.Z. did the numerical simulations.
P.L. derived the DP equations and prepared the figures.
W.Z. made the mathematical proves.
Y.G. collected the simulation/calculation results and wrote the first draft of the manuscript.
P.L., Y.F., X.Z., X.C. and W.Z. wrote and polished the main text and methods section.
X.C. drove the progress of the project.
Z.L, Y.J and H.Y supervised on the project.
All authors reviewed the manuscript.
\end{itemize}

\newpage
\onecolumn
\section*{Methods}
\subsection*{Mathematical description of the conventional Quamtum Readout Error Mitigation}
For an $n$-qubit system, the experimentally measured probability distribution can be treated classically and satisfy
\begin{equation}
    \boldsymbol{P}_{\text{noisy}}=M \boldsymbol{P}_{\text{ideal}},
\end{equation}
where $\boldsymbol{P}_{\text{noisy}}$ ($\boldsymbol{P}_{\text{ideal}}$) is a vector containing $2^n$ outcome probabilities with (without) measurement error and $M$ is a $2^n \times 2^n$ matrix characterizing the most general measurement error.
To characterize $M$, the qubits are selectively flipped and the noisy outcomes are measured under all $2^n$ initial states.
Specially, if the measurement error correlations between different qubits are not considered, the measurement error matrix $M$ takes the form
\begin{equation}
    M=\bigotimes_{i=1}^{n} M_i=\bigotimes_{i=1}^{n} \begin{pmatrix}
        1-\delta_{0,i}&\delta_{1,i}\\\delta_{0,i}&1-\delta_{1,i}
    \end{pmatrix},
\end{equation}
where $M_i$ is the measurement error matrix of each qubit and $\delta_{0,i}$ ($\delta_{1,i}$) is the measurement error rate on qubit $i$ when it is on state $|0\rangle$ ($|1\rangle$).
Meanwhile, the initialization errors can also be modeled as independent and incoherent errors on individual qubits. Since it is in general hard to separate the state preparation and measurement (SPAM) error, they are mixed together in this calibration process.
Then, for each qubit, we use an error mitigation matrix $\Lambda_i$ satisfying
\begin{equation}
    \begin{pmatrix}
        1&0\\0&1
    \end{pmatrix}=\Lambda_i M_i \begin{pmatrix}
        1-q_i &q_i \\ q_i  & 1-q_i
    \end{pmatrix},
\end{equation}
to mitigate the measurement error with the initialization error rate $q_i$ on qubit $Q_i$ \cite{shen_correcting_2012}.
For the whole system, the mitigation matrix $\Lambda$ of the $n$-qubit system can also be written in a tensor form
\begin{equation}
    \begin{aligned}
    \Lambda=\bigotimes_{i=1}^{n} \Lambda_i=\bigotimes_{i=1}^{n} \begin{pmatrix}
        1-q_i &q_i \\ q_i  & 1-q_i
    \end{pmatrix}^{-1}M_i^{-1}=\bigotimes_{i=1}^{n}\frac1{1-2q_i}\begin{pmatrix}1-q_i&-q_i\\ -q_i & 1-q_i
    \end{pmatrix}M_i^{-1}.
    \end{aligned}
\end{equation}
It is noticed that, the mitigation matrix $\Lambda$ is not simply the inverse of measurement error matrix $M$, but also affected by the initialization error of the qubits. In general, if the SPAM error on different qubits are not independent with each other, we can write both the error matrix in the $2^n\times2^n$ matrix form and achieve the correction matrix with $2^n$ measurements.

Here, we use the mitigation process of expectation values $\langle ZZ\rangle$, $\langle ZI\rangle$, and $\langle IZ\rangle$ as examples to show the effect of this QREM process. For an experimentally prepared two-qubit state, the diagonal terms of the density matrix is written as
\begin{equation}
    \boldsymbol{P}_{\text{exp}}=\begin{pmatrix}
        P_{00}\\P_{01}\\P_{10}\\P_{11}
    \end{pmatrix},
\end{equation}
with normalization condition $P_{00}+P_{01}+P_{10}+P_{11}=1$. We can achieve the ideal expectation values
\begin{equation}
\begin{aligned}
\langle ZZ\rangle &=P_{00}-P_{01}-P_{10}+P_{11},\\
\langle ZI\rangle &=P_{00}+P_{01}-P_{10}-P_{11},\\
\langle IZ\rangle &=P_{00}-P_{01}+P_{10}-P_{11}.\\
\end{aligned}
\end{equation}
When measurement error exists, the experimentally measured noisy population vector is
\begin{equation}
    \boldsymbol{P}_{\text{noisy}}=(M_0\otimes M_1) \begin{pmatrix}
        P_{00}\\P_{01}\\P_{10}\\P_{11}
    \end{pmatrix}.
\end{equation}
By applying the QREM matrix, the expectation value is corrected to be
\begin{equation}
\begin{aligned}
    \langle ZZ\rangle_{\text{QREM}}&=\begin{pmatrix}
        1&-1&-1&1
    \end{pmatrix} (\Lambda_0 \otimes \Lambda_1) (M_0\otimes M_1) \begin{pmatrix}
        P_{00}\\P_{01}\\P_{10}\\P_{11}
    \end{pmatrix}\\ &=\frac{\begin{pmatrix}
        1&-1&-1&1
    \end{pmatrix}}{(1-2q_0)(1-2q_1)}\left[{\begin{pmatrix}
        1-q_0 &-q_0 \\ -q_0  & 1-q_0
    \end{pmatrix}}\otimes{\begin{pmatrix}
        1-q_1 &-q_1 \\ -q_1  & 1-q_1
    \end{pmatrix}}\right] (M_0^{-1}\otimes M_1^{-1})(M_0\otimes M_1) \begin{pmatrix}
        P_{00}\\P_{01}\\P_{10}\\P_{11}
    \end{pmatrix}\\
    &=\frac{1}{(1-2q_0)(1-2q_1)}\begin{pmatrix}
        1&-1&-1&1
    \end{pmatrix}\begin{pmatrix}
        P_{00}\\P_{01}\\P_{10}\\P_{11}
    \end{pmatrix}\\
    &=\frac{\langle ZZ \rangle}{(1-2q_0)(1-2q_1)},
\end{aligned}
\end{equation}
\begin{equation}
\begin{aligned}
    \langle ZI\rangle_{\text{QREM}}&=\begin{pmatrix}
        1&1&-1&-1
    \end{pmatrix} (\Lambda_0 \otimes \Lambda_1) (M_0\otimes M_1) \begin{pmatrix}
        P_{00}\\P_{01}\\P_{10}\\P_{11}
    \end{pmatrix}\\ &=\frac{\begin{pmatrix}
        1&1&-1&-1
    \end{pmatrix}}{(1-2q_0)(1-2q_1)}\left[{\begin{pmatrix}
        1-q_0 &-q_0 \\ -q_0  & 1-q_0
    \end{pmatrix}}\otimes{\begin{pmatrix}
        1-q_1 &-q_1 \\ -q_1  & 1-q_1
    \end{pmatrix}}\right] (M_0^{-1}\otimes M_1^{-1})(M_0\otimes M_1) \begin{pmatrix}
        P_{00}\\P_{01}\\P_{10}\\P_{11}
    \end{pmatrix}\\
    &=\frac{1}{1-2q_0}\begin{pmatrix}
        1&1&-1&-1
    \end{pmatrix}\begin{pmatrix}
        P_{00}\\P_{01}\\P_{10}\\P_{11}
    \end{pmatrix}\\
    &=\frac{\langle ZI \rangle}{1-2q_0},
\end{aligned}
\end{equation}
\begin{equation}
\begin{aligned}
    \langle IZ \rangle_{\text{QREM}}&=\begin{pmatrix}
        1&-1&1&-1
    \end{pmatrix} (\Lambda_0 \otimes \Lambda_1) (M_0\otimes M_1) \begin{pmatrix}
        P_{00}\\P_{01}\\P_{10}\\P_{11}
    \end{pmatrix}\\ &=\frac{\begin{pmatrix}
        1&-1&1&-1
    \end{pmatrix}}{(1-2q_0)(1-2q_1)}\left[{\begin{pmatrix}
        1-q_0 &-q_0 \\ -q_0  & 1-q_0
    \end{pmatrix}}\otimes{\begin{pmatrix}
        1-q_1 &-q_1 \\ -q_1  & 1-q_1
    \end{pmatrix}}\right] (M_0^{-1}\otimes M_1^{-1})(M_0\otimes M_1) \begin{pmatrix}
        P_{00}\\P_{01}\\P_{10}\\P_{11}
    \end{pmatrix}\\
    &=\frac{1}{1-2q_1}\begin{pmatrix}
        1&-1&1&-1
    \end{pmatrix}\begin{pmatrix}
        P_{00}\\P_{01}\\P_{10}\\P_{11}
    \end{pmatrix}\\
    &=\frac{\langle IZ \rangle}{1-2q_1}.
\end{aligned}
\end{equation}
The vector $(1,-1,-1,1)^{\rm T}$, $(1,1,-1,-1)^{\rm T}$, and $(1,-1,1,-1)^{\rm T}$ are unnormalized eigenvectors of $(\Lambda_0 \otimes \Lambda_1) (M_0\otimes M_1)$, corresponding to the observables $ZZ$, $ZI$, and $IZ$, respectively.
From the derivation above, it is clear that the QREM process fixes the readout error perfectly and try to correct the initialization error by amplify the result with a factor of $(1-2q_i)^{-1}$ for each measured qubit.

\subsection*{An example of the error introduced by QREM}

Here we use an simple example to demonstrate how entangling gates propagate the initialization error and cause the failure of the conventional QREM method.

We consider the following two-qubit quantum algorithms:\\
$~~~$1.measure the expectation value $\langle ZZ\rangle$ of the initial state $|00\rangle$,\\
$~~~$2.applying a CNOT gate on the initial state $|00\rangle$, and then measure the expectation value $\langle ZZ\rangle$.

When the system is perfectly initialized ($q_0=q_1=0$), for both cases, we can calculate the ideal result with
\begin{equation}
    \langle ZZ\rangle_{ideal}=\begin{pmatrix}
        1&-1&-1&1 
    \end{pmatrix} \begin{pmatrix}
        P_{00}\\P_{01}\\P_{10}\\P_{11}
    \end{pmatrix}=\begin{pmatrix}
        1&-1&-1&1 
    \end{pmatrix} \begin{pmatrix}
        1\\0\\0\\0
    \end{pmatrix}=1.
\end{equation}

In the presence of initialization error, the population vector of the initial state can be written as:
\begin{equation}
    \boldsymbol{P}=\begin{pmatrix}
        (1-q_0)(1-q_1)\\(1-q_0)q_1\\q_0(1-q_1)\\q_0q_1
    \end{pmatrix}.
\end{equation}
In the first case, according to Eq. (S8), we can calculate the QREM corrected result:
\begin{equation}
    \langle ZZ\rangle_{QREM}=\frac{1}{(1-2q_0)(1-2q_1)}\begin{pmatrix}
        1&-1&-1&1
    \end{pmatrix}\begin{pmatrix}
        (1-q_0)(1-q_1)\\(1-q_0)q_1\\q_0(1-q_1)\\q_0q_1
    \end{pmatrix}=1.
\end{equation}
This is exactly the ideal result we expected for. The QREM protocol fixes the initialization error and measurement error simultaneously in this case.
In the second case, when a CNOT gate is applied, the population vector becomes:
\begin{equation}
    \boldsymbol{P}_{exp}=\begin{pmatrix}
        (1-q_0)(1-q_1)\\(1-q_0)q_1\\q_0q_1\\q_0(1-q_1)
    \end{pmatrix}.
\end{equation}
Then, according to Eq. (S8), we can calculate the QREM corrected result:
\begin{equation}
    \langle ZZ\rangle_{QREM}=\frac{1}{(1-2q_0)(1-2q_1)}\begin{pmatrix}
        1&-1&-1&1
    \end{pmatrix}\begin{pmatrix}
        (1-q_0)(1-q_1)\\(1-q_0)q_1\\q_0q_1\\q_0(1-q_1)
    \end{pmatrix}=\frac{1}{1-2q_0}>1.
\end{equation}
From these results, we noticed that the error is related to the initialization error on $Q_0$ (the controlling qubit).
Phenomenologically, comparing Eq. (S14) with Eq. (S12), we found that the CNOT gate swaps the populations on state $|10\rangle$ and $|11\rangle$. When $q_0 \neq 0$, the state of $Q_1$ is thus affected by the initialization error $q_0$. From the above derivation, it is clear that the initialization error can be propagated by the entangling gates and therefore causes the failure of the conventional QREM protocol.

\subsection*{The derivation of entanglement state fidelity overestimation}
To better understand the impact of initialization errors in the preparation of large entangled states,
we hereby reformulate the calculate process using the Pauli basis $\mathcal P^n=\{\langle i\rangle\otimes\{I, X, Y, Z\}\}^{\otimes n}/\langle i\rangle$ ($\langle i\rangle$ is a cyclical group $\{1, i, -1, -i\}$).
In a $n$-qubit system,
the ideal initial state is $|0^{\otimes n}\rangle\langle0^{\otimes n}|=[(I+Z)/2]^{\otimes n}$.
Considering the initialization error,
the initial state $\rho_{\rm ini}$ satisfies
\begin{equation}\label{eq:Ak}
    \rho_{\text{ini}}=\frac1{2^n}\sum_k\lambda_kA_k=\frac1{2^n}\sum_k\lambda_k(Z^{a^0}Z^{a^1}\cdots Z^{a^{n-1}}),
\end{equation}
where $k$ is an integer with its $n$-bit binary representation $k=(a_{n-1}a_{n-2}\cdots a_0)_2=\sum_{i=0}^{n-1}a_i2^i$,
$Z^0=I$,
$Z^1=Z$,
and $A_k\in \mathcal P^n$.
The coefficients $\{a_k\}$ describe classical initial distributions.
For example,
$\forall k\in\{0,1,\cdots,2^n-1\}, a_k=1$ corresponds to the $|0^{\otimes n}\rangle\langle0^{\otimes n}|$.
For the $n$ qubits with independent initialization error rates $q_i$,
\begin{equation}
    \rho_{\text{ini}}=\bigotimes_i\left[(1-q_i)|0\rangle\langle0|+q_i|1\rangle\langle1|\right]=\frac1{2^n}\bigotimes_i\left[(1-q_i)(I+Z_i)+q_i(I-Z_i)\right]=\bigotimes_i\frac{I+(1-2q_i)Z_i}{2},
\end{equation}
therefore $\lambda_k=\prod_i(1-2q_i)^{a_i}$.

In the QREM,
if readout errors and preparation errors are not distinguished,
the corrected result invariably forces $\lambda_k$ to $1$ while simultaneously neglects $\{q_i\}$.
That is,
if the measurement operator is $A_k$,
the final result is tend to multiply the coefficient $1/\lambda_k$ \cite{van_den_berg_model-free_2022}.

We next consider the propagation dynamics of initial state errors under the perfect unitary operation $U$.
\begin{equation}
    U\rho_{\text{ini}}U^\dagger=\frac1{2^n}\sum_k\lambda_kU(Z^{a^0}Z^{a^1}\cdots Z^{a^{n-1}})U^\dagger.
\end{equation}
After a sequence of ideal Clifford operations,
the final state is
\begin{equation}
    \rho_{\text{fin}}=\frac1{2^n}\sum_k\lambda_kS_k,
\end{equation}
where $S_k$ is still Pauli group elements ($S_k\in \mathcal P^n$).
If the sequence of ideal Clifford operations corresponds the graph state (GS) preparation or GHZ state preparation,
$\{S_k\}$ correspond to the state stabilizers one to one.
The circuit preparation scheme governs the one-to-one correspondence of state stabilizers,
such as the linear GHZ (excitation propagates in a chain) or the compact GHZ (excitation propagates from one to the others),
their $\{S_k\}$ are the same set of elements but in different orders.
Meanwhile,
to access the state fidelity,
we measure $\langle S_k\rangle$,
through several single-qubit rotations and the value of $\langle A_{b_k}\rangle$ ($b_k\in\{0, 1, \cdots, 2^n-1\}$).
Thus the actual fidelity $\mathcal F$ and the overestimated $\tilde{\mathcal F}$ are
\begin{equation}\label{eq:clifford_state_fidelity}
    \mathcal F=\frac1{2^n}\sum_k \langle S_k\rangle=\frac1{2^n}\sum_k \langle A_{b_k}\rangle=\frac1{2^n}\sum_k\lambda_k,\text{ and }\tilde{\mathcal F}=\frac1{2^n}\sum_k\frac{\lambda_k}{\lambda_{b_k}},
\end{equation}
affected by initial distribution $\{\lambda_k\}$ and Clifford circuit structure $\{b_k\}$.

Several methods can be employed to calculate the value of $\mathcal F$ and $\tilde{\mathcal F}$.
First,
the effect of Clifford operations on Pauli elements can be efficiently simulated \cite{aaronson2004improved}.
We use binary representation and integer bitwise operations to directly compute the contributions from all stabilizers,
which runs efficiently up to $22$ qubits.
Then,
random sampling of all stabilizers extends to $50-100$ qubits as a rough estimation.
Finally,
we consider a simple case where all qubits have independent and identical initialization error rate, $\forall i, q_i=q$,
and give exact solutions for several typical entangled states.
These solutions are derived by dynamic programming (DP) method within a time complexity of $\mathcal O(4n^2)$,
as following.

\begin{figure*}
    \centering
    \includegraphics[]{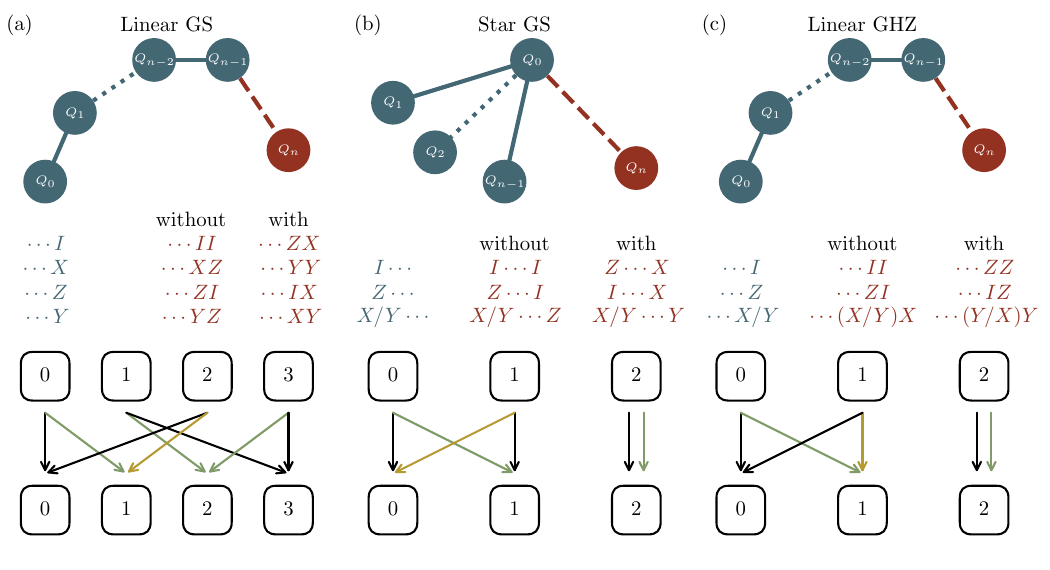}
    \caption{\textbf{Topology structures, state representations, and state transitions of several typical entangling state}.}
    \label{fig:S1}
\end{figure*}

\textbf{The linear graph state}---Considering the $n$ qubit system $Q_0,Q_1,\cdots,Q_{n-1}$ with its topology structure showed in FIG.~\ref{fig:S1}~(a),
the mapping of $\{A_k\}\to\{S_k\}$ is
\begin{equation}
    S_k=\prod_{i=0}^{n-1}a_iS_{2^i},\text{ where }S_{2^i}=\begin{cases}XZII\cdots I,&i=0,\\II\cdots IZXZII\cdots I,&0<i<n-1\text{ with }X\text{ acted on the }Q_i,\\II\cdots IZX,&i=n-1.\end{cases}
\end{equation}
The mapping $\{b_k\}$ is determined by $\{S_k\}$,
Any operation other than the identity $I$ requires a measurement.
Thus, a $Z$ operator is applied to qubit $Q_i$ in the operation $A_{b_k}$,
such as $b_1=3$ with $A_1=ZII\cdots I$, $S_1=XZII\cdots I$, and $A_{b_1}=A_3=ZZII\cdots I$.
In the contribution $\lambda_k / \lambda_{b_k}$, the numerator is given by the number of $Z$ operators in $A_k$, and the denominator by that in $A_{b_k}$, $\lambda_k /\lambda_{b_k}=(1-2q)^{-t}$.
The exponent $t$ characterizes the contribution of stabilizer to the overestimated fidelity $\tilde{\mathcal F}$.

To determine the distribution of the exponent $t$ for all the stabilizers $\{S_k\}$ of $n$ qubits, we first classify $\{S_k\}$ into four categories according to the final Pauli operator, $\{\cdots I\}$ ($s=0$), $\{\cdots X\}$ ($s=1$), $\{\cdots Z\}$ ($s=2$), and $\{\cdots Y\}$ ($s=3$).
We then define a state parameter $d_{n,s,t}\in[0, 1]$, which denotes the fraction of category $s$ contributing $(1-2q)^{-t}$ among all $2^n$ states.
Equivalently, $2^nd_{n,s,t}$ gives the number of stabilizers $\{S_k\}$ that belong to category $s$ and are associated with exponent $t$.
When extending the system to $(n+1)$ qubits,
the state transition equations take the form
\begin{equation}
    \begin{cases}
        2d_{n+1,0,t}=d_{n,0,t}+d_{n,2,t},\\
        2d_{n+1,1,t}=d_{n,0,t-1}+d_{n,2,t+1},\\
        2d_{n+1,2,t}=d_{n,1,t-1}+d_{n,3,t-1},\\
        2d_{n+1,3,t}=d_{n,1,t}+d_{n,3,t}.\\
    \end{cases}
\end{equation}
Then $\tilde{\mathcal F}_n=\sum_{s,t}(1-2q)^{-t}d_{n,s,t}$, with the initial condition $d_{2,0,0}=d_{2,1,0}=1/2$.
The illustration of state transition is showed in FIG.~\ref{fig:S1}~(a).

\textbf{The star-like graph state}---For the mapping $\{A_k\}\to\{S_k\}$ in the star-like GS showed in FIG.~\ref{fig:S1}~(b),
\begin{equation}
    S_{2^i}=\begin{cases}XZZ\cdots Z,&i=0,\\ZII\cdots IXII\cdots I,&0<i\le n-1\text{ with }X\text{ acted on the }Q_i.\end{cases}
\end{equation}
Similar but different from the linear GS,
we classify $\{S_k\}$ into three categories based on the first Pauli operator, $\{I\cdots\}$ ($s=0$), $\{Z\cdots\}$ ($s=1$), and $\{X\cdots,Y\cdots\}$ ($s=2$).
The state transition equations take the form
\begin{equation}
    \begin{cases}
        2d_{n+1,0,t}=d_{n,0,t}+d_{n,1,t+1},\\
        2d_{n+1,1,t}=d_{n,0,t-1}+d_{n-1,1,t},\\
        2d_{n+1,2,t}=d_{n,2,t}+d_{n,2,t-1},\\
    \end{cases}
\end{equation}
with the initial condition $d_{2,0,0}=d_{2,2,0}=1/2$.
The fidelity $\tilde{\mathcal F}_n$ is computed likewise, following a similar procedure.

\textbf{The fully-connected graph state}---For the fully-connect GS,
this special GS has a very strong symmetry,
$S_{2^i}=ZZ\cdots ZXZZ\cdots Z$ with only the $X$ acted on the $Q_i$.
It is noticed that only two type of $\{S_k\}$ is allowed,
\begin{equation}
    S_k=\begin{cases}Y^{a_0}Y^{a_1}\cdots Y^{a_{n-1}},&\text{even number of }Z\text{ in }A_k,\\\left(Z^{1-a_0}X^{a_0}\right)\left(Z^{1-a_1}X^{a_1}\right)\cdots\left(Z^{1-a_{n-1}}X^{a_{n-1}}\right),&\text{odd number of }Z\text{ in }A_k,\\
    \end{cases}
\end{equation}
according to Eq.~(\ref{eq:Ak}).
The latter type contributes the ideal $1$ to the expectation value,
while the former contribute $(1-2q)^{n-t}/(1-2q)^n$ with $(n-t)$ operator $Z$ in $A_k$.
Therefore,
\begin{equation}
    d_{n,t}=\begin{cases}
        \frac12+\frac{1-(-1)^{n-t}}{2^{n+1}}C_n^t,&t=0,\\
        \frac{1-(-1)^{n-t}}{2^{n+1}}C_n^t,&0<t<n,\\
            0,&\text{otherwise},
        \end{cases}
    \end{equation}
which can be derived directly.
The $C_m^n=(m-n)!/(m!n!)$ is the combinatorial number.
The final fidelity $\tilde{\mathcal F}_n=\sum_t(1-2q)^{-t}d_{n,t}$

\textbf{the linear GHZ state}---The quantum circuit for a linear GHZ state generates entanglement sequentially, starting from qubit $Q_0$ and propagating it to each subsequent qubit in the chain.
Thus,
\begin{equation}
    S_{2^i}=\begin{cases}XX\cdots X,&i=0,\\II\cdots IZZII\cdots I,&0<i\le n-1\text{ with }Z\text{ acted on the }Q_{i-1}\text{ and }Q_i.\end{cases}
\end{equation}
At this stage, we classify $\{S_k\}$ into three categories, $\{\cdots I\}$ ($s=0$), $\{\cdots Z\}$ ($s=1$), and $\{\cdots X,\cdots Y\}$ ($s=2$).
Then,
\begin{equation}
    \begin{cases}
        2d_{n+1,0,t}=d_{n,0,t}+d_{n,1,t},\\
        2d_{n+1,1,t}=d_{n,0,t-1}+d_{n,1,t+1},\\
        2d_{n+1,2,t}=d_{n,2,t}+d_{n,2,t-1},\\
    \end{cases}
\end{equation}
with $d_{2,0,0}=d_{2,2,0}=1/2$ and the similar procedure.

\textbf{The "2-cluster tensor" state}---Here we consider a state which is the tensor product of $n/2$ pairs of two-qubit cluster state. Since there is no entanglement between different pairs of cluster state, the measurement outcomes are completely independent, for example:
\begin{equation}
    \langle Z_0X_1X_2Z_3\rangle=\langle Z_0X_1\rangle \langle X_2Z_3\rangle.
\end{equation}
Thus, the estimated fidelity of this state can be calculated through:
\begin{equation}
    \tilde{\mathcal F_n}=\frac{1}{2^n} \sum_{j=0}^{2^n-1} \langle S_{j} \rangle_{\rm QREM}=\prod_{k=0}^{n/2-1}\frac{\langle I_{2k}I_{2k+1}\rangle +\langle Z_{2k}X_{2k+1}\rangle+\langle X_{2k}Z_{2k+1}\rangle+\langle Y_{2k}Y_{2k+1}\rangle}{4}=\prod_{k=0}^{n/2-1}\tilde{\mathcal F_2}=\tilde{\mathcal F_2}^{\frac{n}{2}}.
\end{equation}

\subsection*{The Conditions for the Equalities to hold in Eq.6}\label{prove}
In the main text, we have proved that the fidelity after QREM correction is
\begin{equation}
\begin{split}
\tilde{\mathcal F}&=\frac{1}{2^n} \sum_{j=0}^{2^n-1} \langle S_{j} \rangle_{\rm QREM} \geq \sqrt[2^n]{\prod_{j=0}^{2^n-1} \langle S_j\rangle _{\rm QREM}}\\
&=\sqrt[2^n]{\prod_{i=0}^{n-1} (1-2q_i)^{2^{n-1}-2^{n}(1-f_{i,\mathrm{I}})}} \geq 1.
\end{split}
\end{equation}
The conditions for these two equalities to hold is that $\forall j,j', \langle S_j\rangle_{\rm QREM}=\langle S_{j'}\rangle_{\rm QREM}$ and $\forall i, f_{i,I}=1/2$. Thus, if the fidelity $\tilde{\mathcal F}$ is not over estimated
\begin{equation}
\tilde{\mathcal F}=\frac{1}{2^n} \sum_{j=0}^{2^n-1} \langle S_{j} \rangle_{\rm QREM} = 1.
\label{eq:fidelityeq}
\end{equation}
Together with the equality conditions, we have $\forall j, \langle S_j\rangle_{\rm QREM}=1$. If we consider the stabilizer $S$ which consists only one generator $g_i$ ($S=g_i$). The corrected expectation value is
\begin{equation}
\langle S \rangle=\frac{1-2q_i}{\prod_{k} (1-2q_k)^{c_{i, k}}}=1,
\label{eq:stabilizer}
\end{equation}
where $c_{i,k}=0$ if $g_i$ contains an $I$ on the $k$th qubit, otherwise $c_{i,k}=1$.
Equation~(\ref{eq:stabilizer}) imposes a very strong restriction, in other words, $\sum_{k}c_{i,k}=1$.
If $\forall i, c_{i,k}=\delta_{ik}$, in other words, 
$g_i$ contains only one non-$I$ Pauli operator on the $i$th qubit and contains $I$s on all the other qubits, leading to the Clifford circuit with only single-qubit operations.
Otherwise, the equation $q_i=q_k$ is necessary, while the Clifford circuit also consists of single-qubit operations equivalently with SWAP operations.
No matter which situation it is, the system is thus a tensor product of single qubit states. 

Therefore, we have proved that the state fidelity $\tilde{\mathcal F}$ will not be over estimated only if the system remains in the tensor product state and does not include any entanglement.
Otherwise, when entanglement exists, the state fidelity will always be over estimated after the QREM correction.

\subsection*{Implementation of Variational Quantum EigenSolver and Quantum Time Evolution for Chemistry Systems}\label{vqe}

The electronic Hamiltonian under Born-Oppenheimer approximation has a general form of
	\begin{equation}
		\label{eq-ham}
		H=\sum_{p,q} {h_{pq} a^{\dagger}_{p} a_{q}} + \sum_{p,q,r,s} {\frac{1}{2} h^{pq}_{rs} a^{\dagger}_{p} a^{\dagger}_{q} a_{r} a_{s}},
	\end{equation}
where ${a^{\dagger}_{i}}$ and $a_{j}$ are Fermion creation and annihilation operators, and ${h_{pq}}$ and ${h^{pq}_{rs}}$ in Equation~\ref{eq-ham} refer to one- and two-body integral coefficients. For circuit evolution and measurement, this should be transformed into a linear combination of product of Pauli operators as given below,
\begin{equation}
\label{eq-ham-qubit}
    H = \sum_i c_i P_i, 
\end{equation}
where $P_i$ are tensor products of Pauli operators ($\sigma_x, \sigma_y, \sigma_z, I$).
In the VQE algorithm, the wavefunction is mapping to a parametric quantum circuit, where the ground-state wavefunction and energy satisfy the eigenvalue problem
	\begin{equation}
		H|\Psi\rangle = E|\Psi\rangle
	\end{equation}
In the above framework, the key ingredient is the parametric unitary operator to prepare the wave function ansatz
	\begin{equation}
		\label{eq-general}
		|\Psi(\vec{\theta})\rangle = U(\vec{\theta})|\Psi_{0}\rangle ,
	\end{equation}
where the reference wave function $|\Psi_{0}\rangle$ is usually chosen to be the Hartree-Fock state $|\Psi_{HF}\rangle$.
The parametric wave function is then optimized according to Rayleigh-Ritz variational principle
	\begin{equation}
		\label{eq-vqe}
		E = \min_{\vec{\theta}} {\langle\Psi(\vec{\theta})|H|\Psi(\vec{\theta})\rangle},
	\end{equation}
where the change in parameter values can be calculated on a classical computer using gradient-based or gradient-free optimizers.
The unitary coupled-cluster \cite{ucc-1, ucc-2, ucc-3} (UCC) ansatz is one of the most commonly used physically-motivated ansatz (PMA) in electronic structure simulations. Unlike traditional coupled-cluster theory, which solves a linear amplitude equation, UCC determines the energy and wavefunction variationally via~\ref{eq-vqe}. The unitary operator $U(\vec{\theta})$ is defined as
	\begin{equation}
		\label{eq-ucc}
		|\Psi\rangle = \exp{(T - T^{\dagger})}|\Psi_{0}\rangle,
	\end{equation}
where $|\Psi_{0}\rangle$ is chosen to be the single-determinant Hartree-Fock (HF) wave function. The cluster operator that truncated at single- and double-excitations has the form of
\begin{equation}
  \label{eq::excitation-ops}
		T(\vec{\theta}) = \sum_{p, q} {\theta^{p}_{q} T^{p}_{q}} + \sum_{\substack{p>q\\r>s}} {\theta^{pq}_{rs} T^{pq}_{rs}}
\end{equation}
where the one- and two-body terms are defined as
\begin{equation}
		T^{p}_{q} = a^{\dagger}_{p} a_{q}
\end{equation}
\begin{equation}
		T^{pq}_{rs} = a^{\dagger}_{p} a^{\dagger}_{q} a_{r} a_{s}
\end{equation}
Using fermion-to-qubit transformations such as Jordan-Wigner or Bravyi-Kitaev \cite{jw-bk-1, jw-bk-2, jw-bk-3, jw-bk-4}, the unitary operator $U(\vec{\theta})=\exp(T - T^{\dagger})$ can then be written as:
	\begin{equation}
        U(\vec{\theta})
	    = \exp \left( i\sum_{p, \alpha}{\tilde{\theta}^{\alpha}_{p} \sigma^{\alpha}_{p}} + i\sum_{pq, \alpha\beta}{\tilde{\theta}^{\alpha\beta}_{pq} \sigma^{\alpha}_{p} \sigma^{\beta}_{q}} + \dots \right),
	\end{equation}
where $\{\sigma^{\alpha}_{p}, \sigma^{ \beta}_{q}, \cdots\}$ are Pauli operators $\{ \sigma_{X}, \sigma_{y}, \sigma_{z}, I \}$ on orbitals $\{\alpha, \beta, \cdots, p, q, \dots\}$, and $\{\tilde{\theta}\}$ and $\{\theta\}$ span the same parameter space. These unitary operators are thus decomposed using approximation schemes such as Trotter-Suzuki decomposition \cite{trotter-1, trotter-2} and mapped to quantum circuits, leading to a number of approximately $\mathcal{O}(N^4) \sim \mathcal{O}(N^5)$ gates where $N$ is the number of orbitals.


On a quantum computer, the implementation of the VQE circuit for UCCSD ansatz requires decomposition of the exponential-formed cluster operators into basic quantum single-qubit and two-qubit gates. Approximation schemes are often used, such as Trotter-Suzuki decomposition \cite{trotter-1, trotter-2}:
	\begin{equation}
		\exp(A+B) =  \lim_{N\rightarrow\infty}{\left(e^{\frac{A}{N}} e^{\frac{B}{N}}\right)^{N}}
	\end{equation}.
The Trotterized UCC wave function takes the form:
	\begin{equation}
		|\Psi\rangle = {\prod^{N}_{k=1}{\prod^{M}_{i}{e^{\frac{\theta_{i}}{N}\tau_{i}}}}}|\Psi_{0}\rangle,
	\end{equation}
where $M$ is the total number of individual operators $\tau_{i}$.
For each unitary operator $e^{\frac{\theta_{i}}{N}\tau_{i}}$, a further decomposition is performed using fermion-to-qubit mapping such as Jordan-Wigner or Bravyi-Kitaev:
    \begin{equation}
        e^{\frac{\theta_{i}}{N}\tau_{i}} \rightarrow \prod_{j}{\exp \left(\frac{\tilde{\theta}_{i}}{N} \sigma_{ij}\right)},
    \end{equation}
where $\tau_{i}$ are transformed into linear combinations of Pauli operators $\{\sigma_{ij}\}$. In this way, the exponential of Pauli operators can thus be converted into parametric quantum circuit blocks following \textbf{Algorithm \ref{exp_to_circ}}.
\begin{algorithm}
    \label{exp_to_circ}
    \caption{Map $\exp(i\theta \sigma)$ to a quantum circuit. $\rm HY$ is the Hadamard-Y gate defined as ${\rm HY} = (Z + Y)/\sqrt{2}$}
\label{algo:exp-to-circ}
    \SetAlgoLined
    \KwData{$\sigma$, $\theta$}
    \KwResult{C: the quantum circuit}
    $N_{q}$ $\leftarrow$ number of qubits, C $\leftarrow$ empty circuit\;
    \For{$i=0$; $i\le N_{q}-1$; $i+=1$}
    {
        $p_i$ = $\sigma[i]$\;
        \uIf { $p_i==\sigma_{x}$ }
        {
            C += $H_{i}$
        }
        \ElseIf { $p_i==\sigma_{y}$ }
        {
            C += $\rm{HY}_{i}$
        }
    }
    
    \For{$i=N_{q}-2$; $i \ge 0$; $i-=1$}
    {
        C += ${\rm CNOT}_{(i+1, i)}$
    }
    
    C += ${\rm RZ}(-2\theta)_{N_{q} - 1}$    
    
    \For{$i=0$; $i\le N_{q}-2$; $i+=1$}
    {
        C += ${\rm CNOT}_{i+1, i}$
    }

    \For{$i=0$; $i\le N_{q}-1$; $i+=1$}
    {
        $p_i$ = $\sigma$[i]\;
        \uIf { $p_i==\sigma_{x}$ }
        {
            C += $H_{i}$
        }
        \ElseIf { $p_i == \sigma_{y}$ }
        {
            C += $\rm{HY}_{i}$
        }
    }
\end{algorithm}
The accurate simulation of time evolution under electronic structure Hamiltonians is critical for quantum chemistry applications, particularly in studying non-equilibrium dynamics and spectroscopic properties. Implementing  $U(t) = \exp(iHT)$ on a circuit generally requires Trotterization because terms in $H$ do not commute, as in the UCCSD case introduced before \cite{Babbush_chem_trotter_2015, Hastings_qalgo_chem_2015, Whitfield_qchem_2011}. The Trotter-approximated operator takes the form of
\begin{equation}
\tilde{U}(t) = \prod_{k=1}^{N_{\text{s}}} \prod_i \exp(iP_i \Delta t),
\end{equation}
with $\Delta t = T/N_{\text{s}}$, where higher Trotter steps $N_{\text{s}}$ improves accuracy at the cost of circuit depth. We map Pauli strings to gate sequences using Algorithm \ref{algo:exp-to-circ} described above, yielding HY rotations and CNOT chains.

The VQE benchmarks are carried out for one-dimensional equispaced hydrogen chain. The bond length is fixed as 1.0 \r{A} and canonical orbitals under STO-3G basis set without orbital localization are used to calculate one- and two-electron integrals in molecular orbital basis \cite{pyscf}. Such a configuration lead to the number of qubits being twice of the number of atoms, $n = 2\times n_{\text{atom}}$. Symmetry-reduced UCC ansatz truncated at single and double excitations (sym-UCCSD) \cite{sym_uccsd} is implemented. The circuit which maps exponential Pauli operators to gate sequences is constructed using \textbf{Algorithm \ref{exp_to_circ}}, where parameters are assigned to multiple \textbf{RZ} gates. QREM is carried out inside each VQE iteration. The gradient-free optimizer BOBYQA \cite{pybobyqa} is used for variational optimization.

In the benchmark for quantum time evolution, a hydrogen molecule with bond length r(H-H)=2.0 \r{A} and cc-pVTZ basis set is used under state preparation errors and Trotter approximation. Different active spaces are constructing using the lowest 2 to 16 orbitals, leading to quantum circuits with qubit counts ranging from 4 to 32. The evaluated energy for this systems is
\begin{equation}
E_t = \langle \psi_t | H | \psi_t \rangle,
\end{equation}
where $|\psi_t\rangle \approx \tilde{U}(t) |\psi_{\text{HF}}\rangle$ is generated via Trotterized circuits. The Hartree-Fock energy $E_{0} = \langle \psi_{0} | H | \psi_{0} \rangle$ should provide a theoretical invariant reference that, for perfect evolution $[\exp(iHT), H] = 0$,  the relation $E_t \equiv E_0$ always holds and is independent of time $t$ or the chosen active space.

For both studies, Jordan-Wigner transformation is used to obtain the qubit operators from Fermion excitation operators. Expectation values of Hamiltonian are calculated through tracing the density matrices as $\text{Tr}[\rho H]$ in tensor-network formalism directly, instead of performing measurements then sampling. Initialization errors are implemented by adding Pauli noise channels with given error rate on each qubit at the beginning of the circuit. Electron integrals in molecular orbital basis are calculated using PySCF \cite{pyscf}. Quantum circuit simulations are performed using the Q$^{2}$Chemistry package \cite{q2chem}.





\end{document}